\documentclass[12pt,rmp,twocolumn,natbib,showkeys]{revtex4}

\usepackage[dvips]{graphicx}

\begin{document}

\title{Critical Phenomena and Diffusion in Complex Systems}

\author{\firstname{Bernardo} \surname{Spagnolo$^{\star}$}}
\author{\firstname{Alexander A.} \surname{Dubkov$^{\sharp}$}}
\affiliation{$^{\star}$ Dipartimento di Fisica e Tecnologie Relative
and CNISM-INFM,
\\ Group of Interdisciplinary
Physics\footnote{http://gip.dft.unipa.it}, Universit$\grave{a}$ di
Palermo, \\ Viale delle Scienze, I-90128, Palermo,
Italy\footnote{e-mail: spagnolo@unipa.it}}
\affiliation{$^{\sharp}$
Radiophysics Faculty, Nizhniy Novgorod State University \ \\23
Gagarin Ave., 603950 Nizhniy Novgorod, Russia\footnote{e-mail:
dubkov@rf.unn.ru}}

\begin{abstract}
Editorial of the International Conference on \emph{Critical
Phenomena and Diffusion in Complex Systems} held on 5--7 December,
2006 in Nizhniy Novgorod State University, Russia and was dedicated
to the memory and $80^{th}$ anniversary of Professor Askold N.
Malakhov.
\end{abstract}

\date{\today}

\keywords{Critical phenomena, diffusion phenomena, complex systems}

\maketitle

The nonlinear dynamics of complex systems is one of the most
exciting and fastest growing branches of modern sciences. This
research area is at the forefront in interdisciplinary research and
it has an increasingly important impact on a variety of applied
subjects ranging from the study of turbulence and the behavior of
the weather, through the investigation of electrical and mechanical
oscillations in condensed matter systems, to the physics of
nano-structures and nano-devices, and to the analysis of biological
and economic phenomena. Transport in ion-channels,
synchronization/coherence in biological extended systems, virus
propagation, forecasting protocols are just a few examples that
illustrate the subtle beneficial synergy between noise and
nonlinearity. Complex systems may have extremely rich coherent
dynamics due to the environmental noise and, in specific points of
their phase space, are extremely sensitive to external
perturbations.\\
\indent The performance of any complex system depends on a correct
information exchange between its components. In most natural systems
a signal carrying information is often mixed with noise. Usually the
contamination by the noise makes it difficult to detect signals, but
in some cases noise induced effects known as \emph{stochastic
resonance, resonant activation and noise enhanced stability} improve
conditions for signal detection when noise and system parameters
become ''optimal''~[Gammaitoni \emph{et al.}, 1998; Spagnolo
\emph{et al.}, 2007]. The combined action of external deterministic
or random driving forces and the environmental noise has given rise
to new phenomena with a rich scenario of far-from equilibrium
effects. To describe complex systems it is in fact fundamental to
understand the interplay between noise,
periodic and random driving forces and the intrinsic nonlinearity of the system itself.\\
\indent In-depth study of the multi-fractal space-time structure of
such critical phenomena as seismicity and aftershocks clustering
plays a crucial role in forecasting of seismic activity.
Investigation of such kind of geophysical complex systems, starting
from known empirical laws (Gutenberg--Richter law for earthquakes
magnitudes and heavy tailed Omori law for intensity of aftershocks
triggering) imply developing, in the framework of fractional
diffusion and anomalous branching theory, new general and specific
statistical tools. \\
\indent Anomalous diffusion and fluctuation phenomena in complex
systems were attracting an increasing interest in the last two
decades. A lot of reviews were published and many International
Conferences were organized. Among them we mention the NATO Advanced
Research Workshop ''Stochastic systems: from Randomness to
Complexity'', held in Erice (Italy) in 2002, the International
Workshops ''Noise in Condensed Matter and in Complex Systems''
(2004) and ''Ecological Complex Systems'' (2007), held in Citt\`{a}
del Mare (Palermo, Italy), the ESF--STOCHDYN Conference ''100 Years
of Brownian Motion'', held in Erice (Italy) in 2005, the Centennial
Marian Smoluchowski Symposium held in Krak\'{o}w (Poland) in 2006,
the International Seminar and Workshop ''Constructive Role of Noise
in Complex Systems'', held in Dresden (Germany) in 2006 and the
various reviews concerning  nonlinear fluctuation phenomena and
anomalous diffusion~[Gammaitoni \emph{et al.}, 1998; Metzler \&
Klafter, 2000; Reimann, 2002; Zaslavsky, 2002; Lindner \emph{et
al.}, 2004; see also the focus issue on ''100 Years of Brownian
Motion'',
Chaos \textbf{15} 2005; Chechkin \emph{et al.}, 2006; Spagnolo \emph{et al.}, 2007].\\
\indent The International Workshop "Critical Phenomena and Diffusion
in Complex Systems" was a recent one in the above--mentioned series
of scientific events. This Conference was held on 5--7 December,
2006 in Nizhniy Novgorod State University, Russia and was dedicated
to the memory and $80^{th}$ anniversary of Professor Askold N.
Malakhov (http://www.mptalam.org/conference). Askold N. Malakhov
(1926-2000) was one of outstanding Russian scientists in the field
of statistical radio-physics, Honored Professor of Nizhniy Novgorod
State University, Russian State Prize Laureate, author of four
monographes and more than two hundred scientific papers. His
theoretical and experimental work of fifties and sixties of last
century became the foundation in investigations of the stochastic
behavior of radio--electronic systems with high frequency and time
stability, and influence of the internal noise on the radiation
spectrum of radio--generators. These fundamental results in
statistical radio-physics, important for different applications,
have been published in his first monograph entitled ''Fluctuations
in Self-Oscillating Systems''~[Malakhov, 1968]. This work at once
has received wide recognition among radio-physicists and engineers,
and became the reference book for several generations of experts
working in this area. In seventies the original and highly efficient
methods to solve complex problems of statistical radio-physics were
developed by Askold N. Malakhov in his second monograph ''Cumulant
Analysis of Random non--Gaussian Processes and Their
Transformations''~[Malakhov, 1978]. His further series of papers
regarding the nonlinear random waves became an essential consequence
of his interest to non-Gaussian nonlinear random processes and
fields. The results obtained with the co-authorship of pupils, Prof.
S. N. Gurbatov and Prof. A. I. Saichev, were published in the
book~''Nonlinear Random Waves in Non-dispersive Media'' in
1990~[Gurbatov \emph{et al.}, 1990]. The extended version of this
monograph was later published in England~[Gurbatov \emph{et al.},
1991]. Askold N. Malakhov, in the last years of his life,
successfully developed the theory of diffusive processes in
nonlinear stochastic systems and published some papers in
prestigious journals such as Physical Review
E, Chaos, Advances in Chemical Physics, Technical Physics Letters, etc.\\
\indent The main topics of the International Workshop were
fluctuation phenomena in complex systems, anomalous diffusion and
branching processes, turbulence and nonlinear random waves,
detection of signals and information transfer in nonlinear systems.
The discussion of these problems and those related to the space-time
chaotic behavior of complex self-organizing and multi-fractal
geophysical systems, during the Workshop, has given deep and fresh
insight into stochastic phenomena of different physical nature.\\
\indent About fifty scientists from the USA, Austria, Italy, Poland,
and different Russian Universities (Moscow, Nizhniy Novgorod,
Saratov, Kazan, Ulyanovsk) took part in the International Workshop.
Because of the breadth of Malakhov's scientific interests, and as a
result, of topics of the talks presented at the Conference, it was
organized in two parallel sessions with fourteen invited speakers
and twenty four oral presentations. In the poster session, thirteen
contributions mainly from young researchers, Post-Docs and PhD
students were presented. Most of the presented contributions, after
normal peer review procedure, are collected in
this special issue of \emph{International Journal of Bifurcation and Chaos}.\\
\indent Many presentations have been devoted to diffusion phenomena
in complex systems. Among them we cite the kinetic model of growth
process in crystals, the diffusion of a passive floating impurity in
the one-dimensional channel with molecular collisions and diffusion
of point defects in gallium arsenide, the diffusion of Brownian
particles on a surface with identical potential wells, effects of
thermal fluctuations in the transient dynamics of short and long
Josephson junctions, the stochastic resonance phenomenon in a
mono-stable potentials, noise--induced effects in the polymers
dynamics, fractal branching processes, diffusion in billiards,
anomalous diffusion in one-dimensional fractal Lorentz gas and in
cosmic rays, L\'{e}vy flights in smooth monostable potential
profiles and volatility effects on the escape time in financial market models.\\
\indent From the reports on nonlinear oscillations and waves,
including intensively investigating dynamical chaos, it should be
emphasized the presentations of the well-known Russian scientific
schools of Saratov and Nizhniy Novgorod. They were devoted to the
decay of nonlinear acoustic random waves with non-planar geometry,
chaotic propagation of rays in the underwater acoustics, acoustic
tomography, analysis of transitions to chaos from quasi--periodic
motion on a four-dimensional torus under perturbations by external
noise, synchronization of chaotic self--oscillations, Hodgkin-Huxley
model of a neuron with random excitation, dynamical models of systems with working memory.\\
\indent Among the reports regarding signal processing and
information theory we note the analysis and synthesis of complex
random signals using the third--order spectrum (bispectrum) for
information transfer in parallel communication channels, a new
method of cumulant analysis for quasi--deterministic signals,
bispectral diagnostics in biomedicine, scaling properties of renewal
process with Poissonian statistics, non-Poisson renewal processes,
spectroscopy of flikker--noise based on realizations of a chaotic
signal, analysis of human magneto-encephalograms with the
purpose to reveal photosensitive epilepsy. \\
\indent The reports concerning fluctuations in quantum systems were
devoted to the calculation of the forms--factors of Hydrogen--like
atom nuclei using the effect of Lamb's shift, spectral
characteristics of a two-level quantum system driven by stochastic
fields, nonlinear phenomena of stochastic
quantization.\\
\indent We hopeful believe that the Workshop and this special issue
of \emph{International Journal of Bifurcation and Chaos} will
promote new interdisciplinary studies in the area of critical
phenomena and complex systems. We close this editorial with our
deepest and warmest thanks to all the members of Organizing
Committee and also to all the speakers and participants, whose
discussions, within and after presentations, made the Workshop a
stimulating scientific event. We wish to thank also the referees of
the manuscripts. As a final remark we note that this special issue
gives a widespread view of recent topics and new trends in this rich
and interdisciplinary research field.

\section*{Acknowledgments}

\noindent We acknowledge financial support by Russian Foundation
for Basic Research (Grant No.~06-02-26163).

\vspace{1 cm}
\begin{flushright}
\emph{Bernardo Spagnolo} (Universit\`{a} di Palermo, Italy)
\end{flushright}
\begin{flushright}
\emph{Alexander A. Dubkov} (Nizhniy Novgorod State University,
Russia)
\end{flushright}

\end{document}